\documentclass[aps,prb,preprint,showpacs]{revtex4}
\usepackage{graphicx}

\begin{document}

\title{Electronic Structure of Weakly Correlated Antiferromagnetic
  Metal SrCrO$_3$: First-principles calculations}

\author{Yumin Qian$^1$}
\email{yuminqian@gmail.com}
\author{Guangtao Wang$^2$, Zhi Li$^1$, C. Q. Jin$^1$,
  Zhong Fang$^1$}

\affiliation{$^1$Beijing National Laboratory for Condensed Matter
Physics,and Institute of Physics, Chinese Academy of Sciences,
Beijing 100190, China}
\affiliation{$^2$Department of Physics, Henan
Normal University, Xinxiang 453007, China}

\date{\today}
\pacs{71.20.Be,75.25.Dk,75.50.Ee}

\begin{abstract}
By systematic first-principles calculations, we study the electronic
structure and magnetic property of SrCrO$_3$. Our results suggest that
SrCrO$_3$ is a weakly correlated antiferromagnetic (AF) metal, a very
rare situation in transition-metal oxides. Among various possible AF
states, the C-type spin ordering with small amount of orbital
polarization ($d_{xy}$ orbital is more occupied than the $d_{yz/zx}$
orbital) is favored. The detailed mechanism to stabilize the C-type AF
state is analyzed based on the competition between the itinerant
Stoner instability and superexchange, and our results suggest that the
magnetic instability rather than the orbital or lattice instabilities
plays an important role in this system. The experimentally observed
compressive tetragonal distortion can be naturally explained from the
C-type AF state. By applying the LDA+$U$ method to study this system,
we show that wrong ground state will be obtained if $U$ is large.
\end{abstract}
\maketitle

\section{Introduction}
The electronic and magnetic structures of transition metal perovskites
are usually complicated due to the mutual interplay of various degrees
of freedom~\cite{Imada,Tokura,Dagotto,Fang-1}, such as lattice, spin,
charge and orbital, in the partially filled $d$ shell. The relative
competition or cooperation among those degrees of freedom may
stabilize many possible states, which are energetically close. To
identify those subtle energy differences among various possible
solutions is a hard task and challenging issue for the studies in this
field.  Nevertheless, the first-principles electronic structure
calculations based on density functional theory provides us a
possibility to understand the physical trend and many related issues,
if the studies are done systematically. Such strategy will be followed
in this paper to study the electronic and magnetic properties of
SrCrO$_3$, a simple perovskite but with many controversial issues.

Limited knowledge is available for SrCrO$_3$ due to the requirement
of high pressure for synthesis. Chamberland~\cite{chamber} first
reported SrCrO$_3$ as Pauli paramagnetic metal with cubic perovskite
structure, however in the recent study by Zhou et.al~\cite{Zhou}, it
is suggested as a cubic paramagnetic insulator with a possible
insulator to metal transition under pressure of 4Gpa in their
polycrystal samples. Several anomalous properties, such as deviation
of Curie-Weiss law of magnetic susceptibility, low Seebeck
coefficient, and glassy thermal conductivity, are reported in Zhou's
study, and those properties are attributed to the bond-length
fluctuation instabilities. On the other hand, a new structural
phase, i.e. the perovskite with compressive tetragonal distortion,
was reported recently for SrCrO$_3$ by Attfield
et.al~\cite{attfield2}. A cubic to tetragonal structure transition
was observed around 40K, however no discontinuity of electronic
conductivity was observed due to the coexistence of both cubic and
tetragonal phases even at low temperature. The relative instability
of two structural phases is sensitive to the sample preparation and
microstrain. Attfield et.al~\cite{attfield2} also suggested that, in
the tetragonal phase, the C-type AF state with partial orbital
ordering is favored.  Alario-Franco et.al~\cite{Franco2} again
reported SrCrO$_3$ as a cubic paramagnetic metal, and suggested the
4+ oxidation state of Cr (i.e. Cr$^{4+}$) by electron energy loss
spectroscopy (EELS). No anomaly found in their specific heat
measurement for SrCrO$_3$~\cite{Franco3}.

At the theoretical side, K. W. Lee et.al~\cite{Pick} studied the
electronic structure of SrCrO$_3$ by first-principles calculations.
Their results suggested that the C-type AF spin ordering is the most
stable ground state, and the orbital ordering leads to the tetragonal
lattice distortion. They also found a strong magneto-phonon coupling
for the oxygen octahedral breathing mode and an orbital ordering
transition by LDA+U calculation. Streltsov et.al's~\cite{CaCrO3-1} and
Komarek et.al's~\cite{CaCrO3-2} works focus on CaCrO$_3$, an isovalent
compound with orthorhombic GdFeO$_3$-type distortion. Their results
suggested that CaCrO$_3$ is an intermediately correlated metal with
similar C-type AF ground state, and SrCrO$_3$ is more itinerant than
CaCrO$_3$, making this system especially interesting from both
experimental and theoretical points of view.

In order to understand the electronic and magnetic properties of
SrCrO$_3$, in this work, we perform systematic first-principles
calculations for this compound. The effects of various degrees of
freedom, such as spin, orbital, lattice, and the possible
correlation effect are carefully studied step by step. From the
calculated results, we conclude that the compressive tetragonal
structure with C-type AF spin ordering is the most stable state. A
mechanism is proposed to understand the stabilization of such C-type
AF metallic state. In contrast to earlier studies, we suggest that
the magnetic instability rather than orbital or lattice
instabilities plays main role in this system. The controversial
issues, such as metal or insulator, cubic or tetragonal, can be
naturally clarified based on our systematic analysis. The methods
will be described in section II, and the results and discussions,
which are separated into four subsections step by step, will be
presented in section III. Finally, a brief summary will be given in
section IV.

\section{Method}

The first-principles calculations based on the density functional
theory were performed by using the BSTATE (Beijing Simulation Tool for
Atom Technology) package \cite{bstate}, in which the plane-wave
pseudopotential method is adopted. The 3$d$ states of Cr and 2$p$
states of O are treated with the ultrasoft pseudopotential~\cite{uspp}
and the other states are treated with the optimized norm-conserving
pseudopotential~\cite{norm}. The cut-off energy for describing the
wave functions is 36 Ry, while that for the augmentation charge is 200
Ry. We use $(8\times8\times8)$ mesh for the k-points sampling in the
linear tetrahedron method with the curvature correction. For the
exchange-correlation energy, the local density approximation (LDA) is
used. We further use the LDA+$U$ method to study the effect of
electron correlation~\cite{bstate}. Virtual crystal
approximation(VCA)~\cite{VCA} is used to study the effect of doping.

Both cubic and tetragonal perovskite structures with full lattice
relaxation are studied . For each fixed crystal structure, various
different magnetic states are studied. They are, nonmagnetic (NM)
state, ferromagnetic (FM) state, layered-type antiferromagnetic
(A-type AF) state, chain-type antiferromagnetic (C-type AF) state, and
the conventional antiferromagnetic (G-type AF)
state~\cite{Imada}. Local atomic orbitals of Cr atom are defined to
calculate the orbital occupation and its ordering.

\section{Results and Discussions}

In order to understand the physics clearly, it is better to separate
the contributions coming from different degrees of freedom. We
therefore, in this study, use the following strategy: starting from
the simplest situation, where various degrees of freedom are frozen
(not involved), then we add the effect of different degrees of freedom
one by one to see the evolution of electron structure under various
conditions.

\subsection{Basic Electronic Structure of NM SrCrO$_3$ in Cubic Structure}

\begin{figure}[tbp]
\includegraphics[clip,width=8cm]{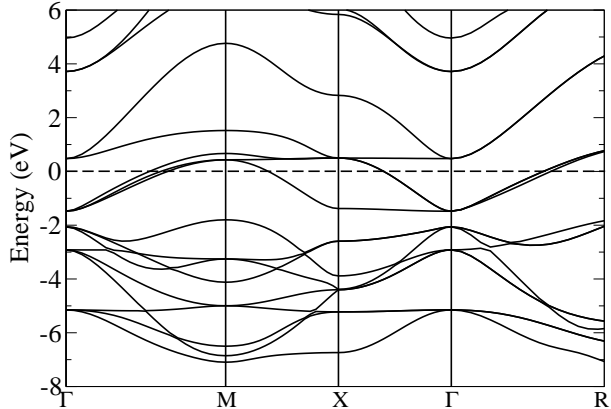}
\caption{\label{NMband}Band structure of NM SrCrO$_3$ in cubic
  structure along high symmetry lines. The Fermi level $E_{F}$ is at
  0.0eV}
\end{figure}

\begin{figure}[tbp]
\includegraphics[clip,width=8cm]{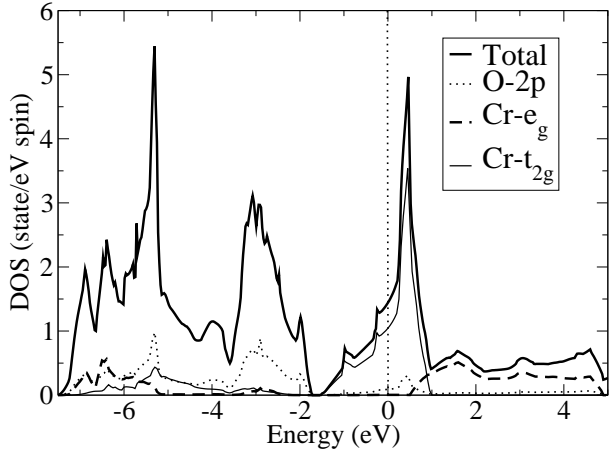}
\caption{\label{NMPDOS}The density of states (DOS) and projected DOS
  (PDOS) of NM SrCrO$_3$ in the cubic perovskite structure. The Fermi
  level is at energy zero. A sharp DOS peak around 0.5eV is due to
  the flat band segments visible in Fig\ref{NMband}.}
\end{figure}

The nominal atomic valences in SrCrO$_3$ are Sr$^{2+}$, Cr$^{4+}$
and O$^{2-}$, respectively, where the $2p$ states of O$^{2-}$ are
fully occupied, and the $3d$ shell of Cr$^{4+}$ is filled by two
electrons, similar to the $d^2$ systems like LaVO$_3$ or
YVO$_3$~\cite{LVO1,LVO2}or CrO$_2$~\cite{CrO2}. Since the atomic
levels of Sr$^{2+}$ are far away from the Fermi level, the
electronic properties of SrCrO$_3$ are mostly determined by the
$p$-$d$ bonding between Cr-$3d$ and O-$2p$. Fig.\ref{NMband} and
Fig.\ref{NMPDOS} show the electronic band structure and density of
states (DOS) of NM SrCrO$_3$ in the cubic perovskite structure,
calculated with experimental lattice parameter (space group $Pm3m$,
a=3.811\AA)~\cite{attfield2}. The states from -7.4eV to -1.6eV are
mostly from the O-$2p$ bonding orbitals, and the states around the
Fermi level $E_f$ (from -1.5eV to 5.0eV) are from Cr-$3d$
antibonding orbitals, as shown in the projected DOS (PDOS)
Fig\ref{NMPDOS}. The $p$-$d$ hybridization is substantial, and the
total DOS at the Fermi level is about 1.48 (States/eV spin f.u.).

Due to the octahedral local crystal field, the five $d$ orbitals of Cr
split into a lower lying three-fold degenerate $t_{2g}$ manifold and a
higher two-fold degenerate $e_g$ manifold. The band width of $t_{2g}$
and $e_g$ states are about 2.5eV and 4.2eV respectively.  The $e_g$
band width is much larger than that of $t_{2g}$, because the
$pd\sigma$-type bonding is stronger than $pd\pi$-type bonding. The
$t_{2g}$-$e_g$ separation (crystal field splitting) is about 2.4eV,
comparable to the $t_{2g}$ band width. The Fermi level $E_{f}$ lies
within the $t_{2g}$ manifold, while the $e_g$ part lies well above
$E_{f}$, suggesting a typical $t_{2g}$ system. The calculated total
occupation number of Cr-$3d$ orbitals is $n=1.994$, quite consistent
with chemical stoichiometry analysis and the EELS experiment which
suggests the $Cr^{4+}$ valence state by Alario-Franco\cite{Franco3}.

\begin{figure}[tbp]
\includegraphics[clip,width=2.4in,height=1.8in]{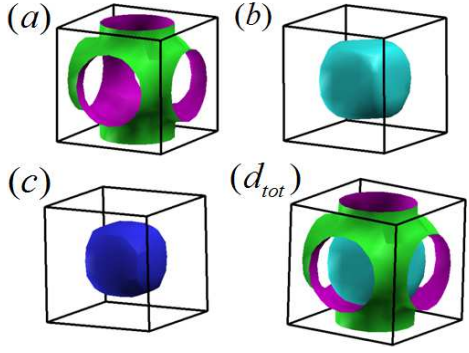}
\caption{\label{FS}The Fermi Surfaces (FS) of NM cubic
  SrCrO$_3$. (a-c) The three sheets of FS, and ($d_{tot}$) the display
  for total three sheets. The $\Gamma$-point locates at the center
  of the cube.}
\end{figure}

\begin{figure}[tbp]
\includegraphics[clip,width=8cm]{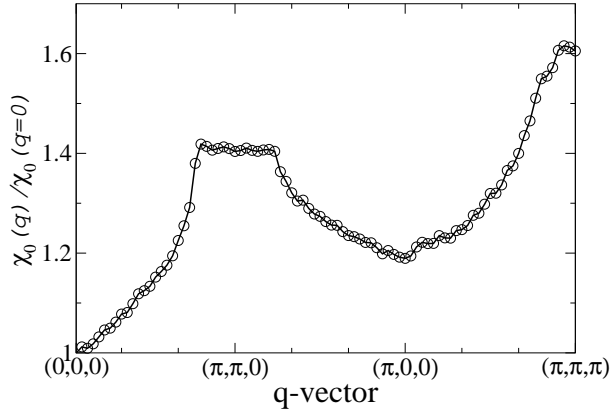}
\caption{\label{nesting} The calculated Lindhard response function
  $\chi_{0}(q)$ for $q$-vectors along high symmetry lines
  $\Gamma$-M-X-R of NM cubic SrCrO$_3$. There are two structures
  around $q=(\pi,\pi,0)$ and $q=(\pi,\pi,\pi)$, respectively,
  corresponding to two different kinds of possible instabilities. }
\end{figure}

All the three $t_{2g}$ bands cross the Fermi Level, forming three
sheets of Fermi surface (FS), as shown in Fig.\ref{FS}. There is
also a large section of flat band structure along $\Gamma
(0,0,0)-X(0,\pi,0)-M(\pi,\pi,0)$ high symmetry lines, which leads to
the sharp DOS peak located at $0.5$eV above $E_f$. The existence of
such Van-Hove singularity will affect the electronic structures
considerably, as will be addressed below.  Among the three sheets of
FS, two of them are cubic like with flat facets, suggesting the
possible existence of FS nesting~\cite{Pick}. In order to clarify
this point, we calculated the Lindhard response function
$\chi_0(q)$, as shown in Fig.\ref{nesting}. There are a plateau
locating around $q=(\pi,\pi,0)$ and a sharp peak locating at
$q=(\pi,\pi,\pi)$, with the latter peak slightly higher than the
former. The presence of two structures in the response function
corresponds to two kind of instabilities, which are related to the
C-type and G-type AF ordering respectively, as will be discussed
below. Considering the fact that the peak at $q=(\pi,\pi,\pi)$ is
higher than that at $q=(\pi,\pi,0)$, it may be expected that the
G-type AF state is easier to be stabilized than the C-type AF state.
Unfortunately, this expectation is not consistent with our following
calculations.

\subsection{Magnetic Instability in Cubic Structure}

\begin{figure}[tbp]
\includegraphics[clip,width=8cm]{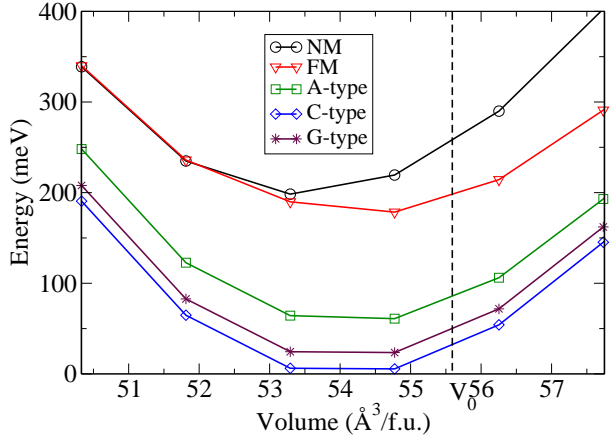}
\caption{\label{EV_LDAc}The calculated total energy versus volume for
  cubic SrCrO$_3$ in various spin states. $V_{0}$ is the
  experimental volume.}
\end{figure}

\begin{figure}[tbp]
\includegraphics[clip,width=8cm]{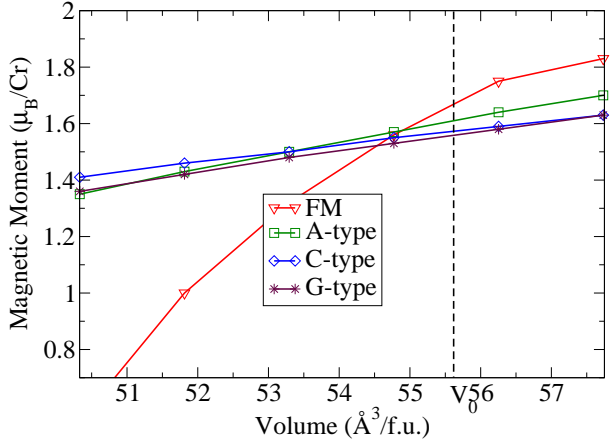}
\caption{\label{mag_vol} The calculated volume dependency of magnetic
  moment for various spin states of cubic SrCrO$_3$. $V_{0}$ is the
  experimental volume.}
\end{figure}

\begin{figure}[tbp]
\includegraphics[clip,width=8cm]{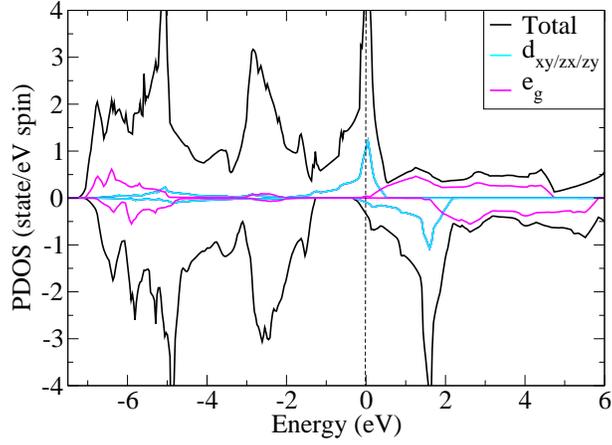}
\caption{\label{PDOS_FM} Projected density of state(PDOS) of FM spin
state in cubic structure.}
\end{figure}

\begin{figure}[tbp]
\includegraphics[clip,width=8cm]{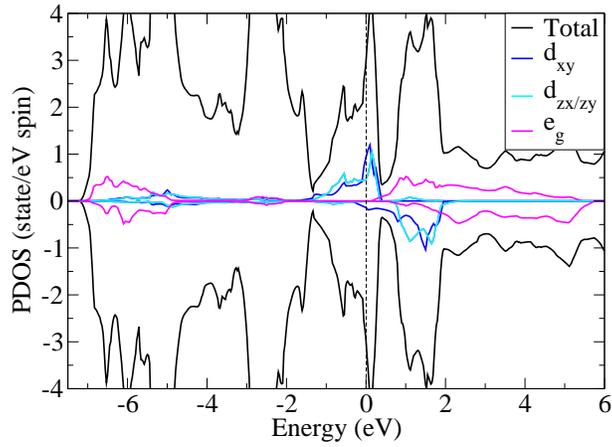}
\caption{\label{PDOS_a} Projected density of state(PDOS) of A-type AF
  state in cubic structure.}
\end{figure}

\begin{figure}[tbp]
\includegraphics[clip,width=8cm]{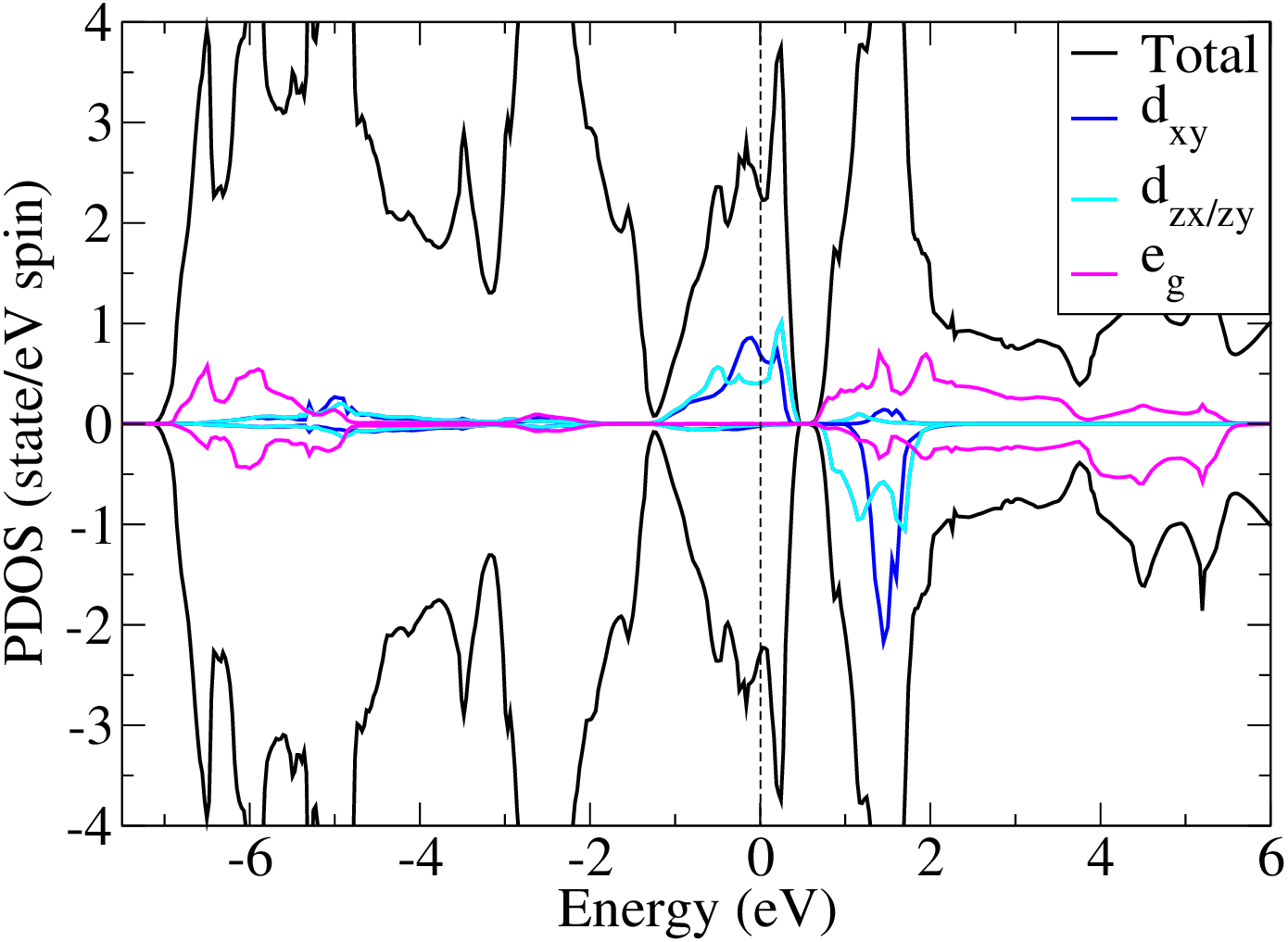}
\caption{\label{PDOS_c} Projected density of state(PDOS) of C-type AF
  state in cubic structure.}
\end{figure}

\begin{figure}[tbp]
\includegraphics[clip,width=8cm]{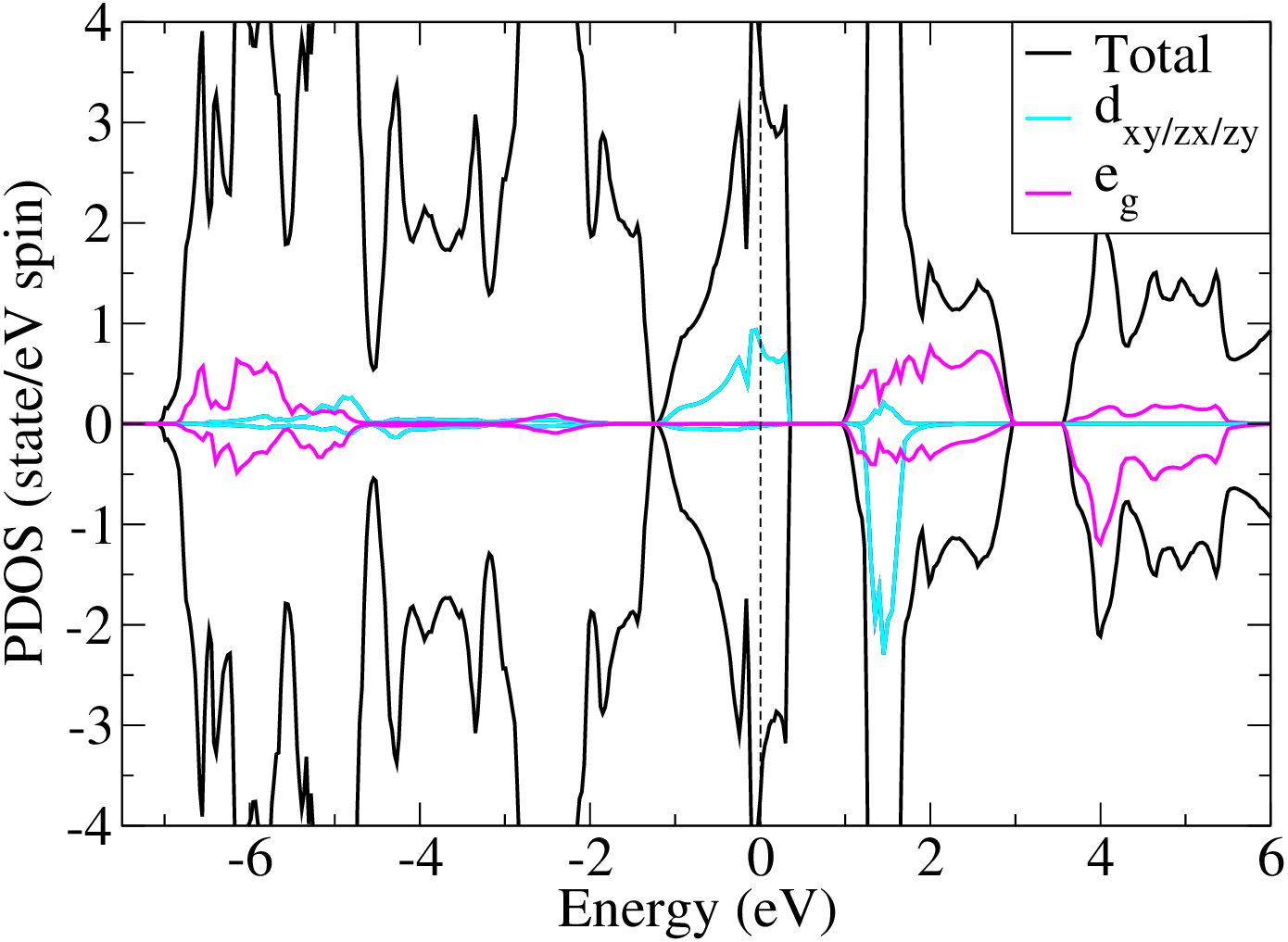}
\caption{\label{PDOS_G} Projected density of state(PDOS) of G-type AF
  state in cubic structure.}
\end{figure}

Now we start to include the spin degree of freedom into our studies.
For this purpose, we neglect the possible lattice distortion, and
only concentrate on the cubic structure.  We calculate the total
energies of different magnetic states as function of volume. As
shown in Fig.\ref{EV_LDAc}, the C-type AF state is always stabilized
for all the calculated volumes. In other words, the C-type AF state
is the most stable state even without any lattice distortion. This
is one of our main conclusions, which will be carefully discussed in
this part.

From the energetic point of view, as shown in Fig.\ref{EV_LDAc}, the
energy difference between the NM and the FM states is small (about
54meV/f.u., if experimental lattice parameter is used), and the FM
state is easily suppressed by volume compression. On the other hand,
there exists a big energy difference (larger than 100meV/f.u.)
between the FM and various AF states, and the energy differences
among various AF states (A-type, C-type and G-type) are small (about
20meV/f.u.$\sim$50meV/f.u.). In view of this fact together with the
fact suggested by calculated response function Fig.\ref{nesting}
that itinerant picture alone can not fully explain the magnetic
property, the superexchange must play some role to stabilize the AF
states. The subtle balance between the itinerant kinetic energy and
superexchange leads to the C-type AF ground state.

The calculated magnetic moments of various states as functions of
volume are shown in Fig.\ref{mag_vol}. At the equilibrium volume, the
high spin states are always favored, and the magnetic moments of
various states are close to 2.0$\mu_B$/Cr, suggesting the sizable
Hund's coupling. The calculated moment are 1.63$\mu_B$/Cr, 1.59$
\mu_B$/Cr, 1.56$\mu_B$/Cr, 1.55$\mu_B$/Cr for FM, A-type, C-type, G-type
states respectively. For the FM state, the magnetic moment quickly
collapses with volume compression as show in Fig.\ref{mag_vol}, but
for the AF states (A-type, C-type and G-type), the moments are not so
sensitive to the volume change.  This again suggests that although the
system is close to itinerant Stoner instability, the stabilization of
AF states is beyond the Stoner mechanism.

Although we only treat the cubic structure, the A-type and C-type
magnetic ordering will further break the symmetry, and lead to orbital
ordering. As shown in Fig.\ref{PDOS_FM}-\ref{PDOS_G} and listed in
Table I, the degeneracy between the $d_{xy}$ and the $d_{yz/zx}$
orbitals is lifted in the presence of A-type or C-type AF states. More
importantly, the A-type and C-type AF states lead to different orbital
ordering: the occupation of $d_{xy}$ orbital is larger (smaller) than
$d_{yz/zx}$ orbitals for the C-type (A-type) AF state. Recall the fact
that current calculations are based on cubic structure, the observed
orbital ordering must be purely due to the effect of magnetic
orderings. If we evaluate the orbital polarization as the occupation
number difference between the $d_{xy}$ and the $d_{yz/zx}$ orbitals,
defined as $n_p=n_{d_{xy}}-n_{d_{yz/zx}}$, the calculated values
suggest that the polarization is about $n_p=-0.04$ and $n_p=0.06$
electrons for the A-type and C-type AF state respectively.

Now we have to answer several important questions: (1) why the types
of orbital ordering are opposite for A-type and C-type AF states?
(2) why the C-type AF state is more stable than the A-type AF state?
To answer these questions, the detailed understanding to the
superexchange is important. Through the superexchange mechanism, the
system gains energy from hybridizations: the hybridization between
the occupied states (at one site) and the unoccupied states (at
neighboring sites) will lower the energy of the occupied states by
approximately $t^2/\Delta$, where $t$ is an effective transfer
integral and $\Delta$ is an energy difference between occupied and
unoccupied states. We start from the C-type AF state in which the
spins of two adjacent Cr$^{4+}$ ions lie antiparallely in $ab$-plane
but parallely along $c$-axis. The occupied majority spin $d_{xy}$
($d_{yz/zx}$) orbital is pushed down by hybridization with
unoccupied majority spin $d_{xy}$ ($d_{yz/zx}$) orbital located at
nearest-neighboring Cr site in $ab$-plane (The effect is forbidden
along the $c$-axis due to the parallel spin chain). Such
hybridization process can happen four times for $d_{xy}$ orbital,
but only two times for $d_{yz/zx}$ orbital due to its spacial
orientation. Therefore, the occupied majority spin $d_{xy}$ orbital
is pushed down more strongly than the $d_{yz/zx}$ orbitals due to
the larger hybridization. Applying the same analysis to the A-type
AF state in which the spins of two adjacent Cr$^{4+}$ ions lie
parallely in $ab$-plane but antiparallely along $c$-axis, only the
hybridization paths along the $c$-axis need to be considered. Now,
for the A-type AF state, the hybridization process can happen two
times for $d_{yz/zx}$ orbital, but zero time for $d_{xy}$ orbital
again due to its spacial orientation.  Therefore, the $d_{yz/zx}$
state is lower than the $d_{xy}$ state for the A-type AF state, in
opposite to the situation in C-type AF state. This will answer the
first question listed above. Due to the same reason, the system can
gain energy from eight possible hybridization paths in the C-type AF
state, but only four paths in A-type AF state. This is the reason
why the C-type AF state is more stable than A-type AF state.

According to the above mechanism, it is immediately realized that
the G-type AF state should be more stable than A-type or C-type AF
states. Why it is not the case from the first principle
calculations? The reason is following: applying the superexchange
mechanism in the above discussion, we assume that the majority spin
$d_{xy}$ or $d_{yz/zx}$ orbitals are fully occupied. This is of
course not the case, due to the large band width and the small
splitting between these $t_{2g}$ orbitals, the system is always
metallic and these $t_{2g}$ states are partially occupied. In this
situation, Stoner mechanism, which favors itinerant ferromagnetism
due to the kinetic energy gain, also plays its role, and finally the
full AF state (i.e, the G-type AF state) lost its energy gain. The
system prefers to keep certain FM spin orientation along particular
directions, such as $c$-axis in C-type and $ab$-plane in A-type AF
state. Therefore, the stabilization of C-type AF state comes from
the subtle balance between the itinerant Stoner mechanism and the
superexchange.

\begin{figure}[tbp]
\includegraphics[clip,width=8cm]{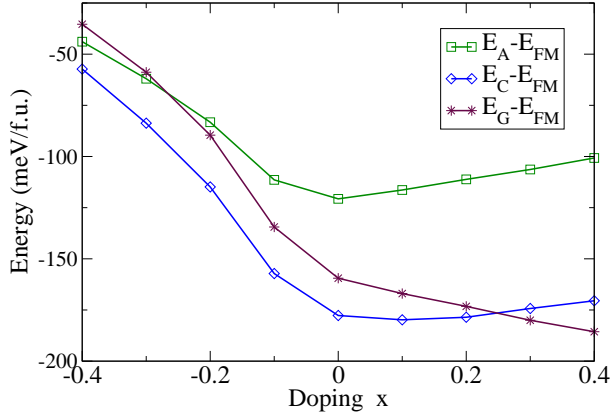}
\caption{\label{E-doping} The calculated total energy (relative to
the FM spin state) as function of charge doping by Virtual Crystal
Approximation(VCA).}
\end{figure}

To further prove our above discussion, we performed the following
calculations. Keeping the cubic lattice structure, we calculate the
energy difference between various magnetic state as function of number
of electrons. The virtual crystal approximation(VCA)~\cite{VCA} is used
here to simplify our studies.  For the parent compound, the total
number of $t_{2g}$ electron on Cr$^{4+}$ is 2. By increasing the
number of $d$ electrons, the majority spin $t_{2g}$ states tend to be
fully occupied.  As we can see from Fig.\ref{E-doping}, if we increase
the number of electrons, indeed the G-type AF state is stabilized.

\subsection{Effect of Lattice Distortion}

In our above analysis, only the cubic structure is treated. However,
the tetragonally distorted structure is also observed experimentally,
the possible tetragonal distortion and its effect on electronic
structure must be studied.  The tetragonal phase must be energetically
very close to the cubic phase, duo to the reported coexistence of
these two phases at low temperature\cite{attfield2}.  Therefore, in
this section, we try to understand the effect of tetragonal structure
distortion. The main conclusions drew from this section are following:
(1) If tetragonal distortion is allowed, the compression (rather than
elongation) along the $c$-axis should be favored, in consistent with
experimental observation. (2) C-type AF state and corresponding
orbital ordering are further stabilized by compressive tetragonal
distortion. (3) The energy gain due to tetragonal distortion (about
10meV/Cr) is smaller than that coming from the superexchange (about
100meV/Cr) discussed above. (4) The orbital polarization induced by
tetragonal distortion is much smaller than that caused by magnetic
orderings.

\begin{figure}[tbp]
\includegraphics[clip,width=8cm]{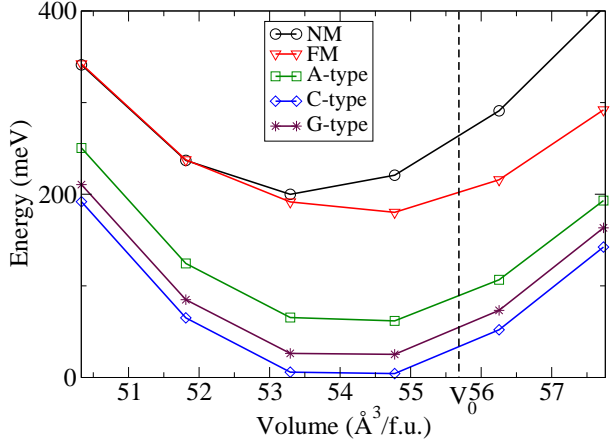}
\caption{\label{EV_LDAt}The calculated total energy versus volume for
  all kinds of spin states by LDA. The lattice distortion (c/a ratio)
  is optimized for each fixed volume, and $V_{0}$ is the experimental
  volume.}
\end{figure}

\begin{figure}[tbp]
\includegraphics[clip,width=8cm]{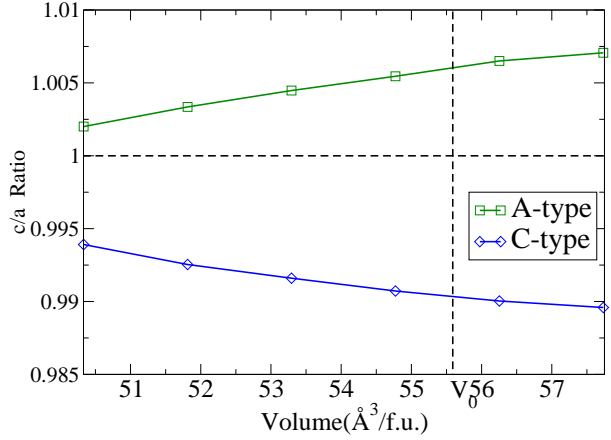}
\caption{\label{optim}The optimized $c/a$ ratio for A-type and C-type
  AF state at different volume. Volume compression will suppress the
  lattice distortion. $V_{0}$ is the experimental volume.}
\end{figure}

Fig.\ref{EV_LDAt} shows the optimized total energy as function of
volume after allowing the tetragonal distortion. Comparing with the
cubic results shown in Fig.\ref{EV_LDAc}, the qualitative picture is
not changed at all. The C-type AF state is still the most stable
state, and a big energy difference exists between the FM state and the
various AF states. However, what is interesting is that the C-type AF
state favors compressive distortion, and A-type AF state favors
elongation distortion along the $c$-axis as shown in
Fig.\ref{optim}. For the most stable C-type AF state, the optimized
c/a ratio is 0.990, in good agreement with the experimentally observed
tetragonal distortion (c/a=0.992). Such an opposite tendency for
A-type and C-type AF states can be easily understood from the orbital
polarization discussed above. In the C-type AF state, the $d_{xy}$
orbital is more occupied than the $d_{yz/zx}$ orbitals, therefore
$c$-axis compression is favored due to the reduction of electron
static potential. The opposite is true for A-type AF state. As we
expected, the lattice distortion will further enhance the orbital
polarization, and stabilize the A or C-type AF states. In reality,
A-type and C-type AF states are further stabilized by only $2.0$meV/Cr
and $6.0$meV/Cr respectively, comparing with their cubic phase. On the
other hand, the FM and G-type AF states have no energy gain by lattice
distortion.

Another important issue is the origin of orbital polarization observed
both in experiment and LDA calculations. Since both lattice distortion
(due to crystal field) and AF magnetic ordering due to superexchange
mechanism can lead to orbital polarization, we have to determine which
one is the dominate factor. From our LDA calculations shown in table
I, it is suggested that the magnetic ordering should be the dominant
factor. First of all, the energy gain due to the lattice distortion is
small (less than 10meV), even smaller than the energy difference among
the A-type, C-type, G-type AF states (about$20\sim50$meV ) calculated
in the cubic phase. Second, as shown in table I, where the calculated
orbital polarization of various magnetic states for different $c/a$
ratio are listed, the orbital polarization coming from the lattice
distortion is smaller than that coming from the superexchange effect
calculated in the cubic phase.

Finally, we conclude that although the tetragonal distortion is
favored, the distorted phase is energetically close to the cubic
phase, in consistent with experimental results of coexistence of the
two phases. On the other hand, an electronic structure transition
reported experimentally around 4 GPa~\cite{Zhou} is not observed from
our calculations.

\begin{table}[htbp]
\caption[]{The orbital polarization, evaluated as the occupation
  number difference between the $d_{xy}$ and the $d_{yz/zx}$ orbitals
  (i.e, $n_p=n_{d_{xy}}-n_{d_{yz/zx}}$), for various magnetic states
  with different $c/a$ ratio.} \label{tab:orbit} \vspace{4mm}
\begin{tabular}{|c|c|c|c|c|c|c|c|}
\hline
              & \multicolumn{7}{|c|} { $c/a$ Ratio}\\
\cline{2-8}
                & 1.015 &1.010 &1.005 &1.000 &0.995 &0.990 &0.985\\
\cline{1-8}
      FM        & -0.041 &-0.028 &-0.014 & 0.000 & 0.015 & 0.028 & 0.042\\
      A-type    & -0.061 &-0.053 &-0.046 &-0.038 &-0.031 &-0.023 &-0.011\\
      C-type    &  0.022 & 0.033 & 0.043 & 0.056 & 0.067 & 0.076 & 0.087\\
      G-type    & -0.031 &-0.019 &-0.009 & 0.000 & 0.011 & 0.021 & 0.032\\\hline
\end{tabular}
\end{table}

\subsection{The Effect of On-site Coulomb Correlation}

\begin{figure}[btp]
\includegraphics[clip,width=8cm]{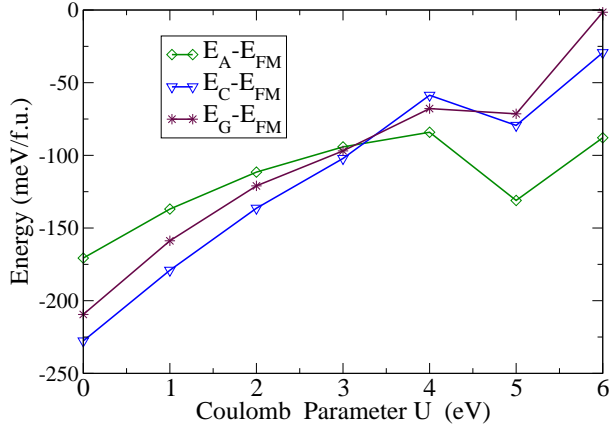}
\caption{\label{E_U}The calculated total energy (relative to the FM
  solution) of different magnetic states as function of $U$ by
  LDA+$U$ method.}
\end{figure}

\begin{figure}
\includegraphics[clip,width=8cm]{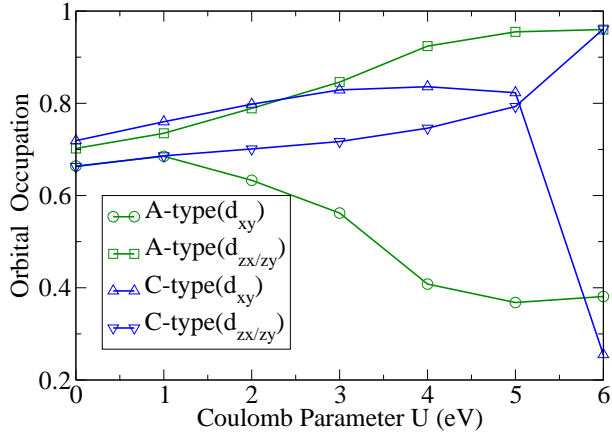}
\caption{\label{OO_U} Orbital occupation versus correlation parameter
  $U$ for A-type and C-type AF state in the cubic lattice structure
  calculated by LDA+$U$.}
\end{figure}

One important issue in transition metal compounds is the possible
strong correlation, which can not be treated successfully by LDA
level calculations. We therefore need to check the effect of
electron correlation by advanced technique. Given the fact that
SrCrO$_3$ should be more itinerant than CaCrO$_3$ due to the lager
band width, and also the fact that most of the electron properties
calculated above using LDA for SrCrO$_3$ can be well compared with
experiment, we expect that the effect of on-site Coulomb $U$ is not
significant. Nevertheless, we will show, in this part by LDA+$U$
calculations, that the SrCrO$_3$ system is indeed not strongly
correlated system. In principle, LDA+$U$ is a cheap technique, which
is suitable for strongly correlated AF insulating state (large $U$
limit), however we will show that the LDA+$U$ results for SrCrO$_3$
are nonphysical if large $U$ is applied.

Fig.\ref{E_U} shows the calculated total energy relative to FM state
as function of $U$. We can see clearly that the A-type AF state is
stabilized at large $U$ limit ($U>4.0$eV), which is not consistent
with experiment. What is more surprising is that, for the C-type AF
state, an orbital ordering transition is observed, as shown in
Fig.\ref{OO_U}. For the small $U$ limit, the $d_{xy}$ orbital is more
populated than $d_{yz/zx}$ orbitals, however for the large $U$ limit,
the $d_{yz/zx}$ orbital become more occupied, in opposite to the small
$U$ limit. The appearance of this transition can be understood as the
following. In the high spin configuration of Cr$^{4+}$, the two $3d$
electrons will occupy the $t_{2g}$ orbitals, which split into one
$d_{xy}$ orbital and two degenerate $d_{yz/zx}$ orbital in the C-type
AF state. If the $d_{xy}$ orbital is lower in energy, the system has
to be metallic due to the degeneracy between the $d_{yz}$ and the
$d_{zx}$ orbitals. However, if the $U$ is very large (large than band
width), it will try to reverse the energy level ordering of $d_{xy}$
and $d_{yz}/d_{zx}$ states, in such a way that two $d_{yz/zx}$
orbitals are mostly occupied and $d_{xy}$ becomes nearly empty, then
the system gains energy from the possible gap opening between the
occupied $d_{yz/zx}$ and unoccupied $d_{xy}$ orbitals if $U$ is large.

The above analysis for orbital polarization suggest that, no matter
the system is ordered in the C-type or the A-type AF states, the
$d_{yz/zx}$ orbitals are always more favored than the $d_{xy}$ orbital
in the large $U$ limit. In such an orbital polarization, the
elongation (rather than compressive) tetragonal distortion should be
favored, in opposite to experimental observation. Therefore, our
calculations support the conclusion that the SrCrO$_3$ is at least not
strongly correlated system. Furthermore, from above studies, we also
learn that the metallicity of SrCrO$_3$ in the C-type AF state is
protected by the degeneracy between the $d_{yz}$ and the $d_{zx}$
orbitals given the compressive tetragonal distortion observed.  From
the very good agreement between the optimized $c/a$ ratio from LDA
calculation (c/a=0.990) and that observed experimentally (c/a=0.992),
we can further conclude that the effect of correlation is in fact very
weak.

\section{Summary}

In summary, we study the electronic structure and the magnetic
property of SrCrO$_3$ systematically based on the LDA and LDA+$U$
calculations. We analyze the various magnetic instabilities, the
effect of lattice distortion, and the effect of correlation. The
controversial issues, such as metal or insulator, cubic or
tetragonal, can be now understood in a consistent picture.  Our
calculated results are also quite consistent with available
experiment results. We finally conclude that SrCrO$_3$ is a weakly
correlated AF metal with small amount of orbital polarization. The
magnetic instability rather than the orbital or lattice
instabilities plays the dominant role in this compound.

\section{Acknowledgments}
We acknowledge the valuable discussions with H. J. Zhang, X. Y. Deng,
and the supports from the NSF of China, the 973 Program of China, and
the International Science and Technology Cooperation Program of China.

\end{document}